\documentstyle[aps,prd,floats,epsf]{revtex}
\begin{document}
\draft % Makes pacs numbers print

%%%%%%%%%%%%%%%%%%%%%%%%%%%%%%%
% Shortcut definitions
%%%%%%%%%%%%%%%%%%%%%%%%%%%%%%%
\def\dt{{\hat \partial}_0}
\def\dr{{\partial\over\partial r}}

\def\gamr{{\Gamma^r_{rr}}}%\def\gamr{{\Gamma^r{}_{rr}}}
\def\dgamr{{\partial{\Gamma^r{}_{rr}}\over\partial r}}
\def\grt{{\Gamma_{rT}}}
\def\mrr{{M_{rrr}}}
\def\mtr{{M_{Tr}}}
\def\mrt{{M_{rT}}}
\def\kr{{K_{rr}}}
\def\kt{{K_T}}
\def\lr{{L_{rr}}}
\def\lt{{L_T}}
\def\ar{{a_r}}
\def\arr{{a_{rr}}}
\def\at{{a_T}}
\def\aor{{a_{0r}}}
\def\krtil{{K^r_r}}%\def\krtil{{K^r{}_r}}
%%%%%%%%%%%%%%%%%%%%%%%%%%%%%%%

%--------------------------------------------------------------------
%--------------------------------------------------------------------
%  This is the first line to be uncommented for 2 column format
\twocolumn[\hsize\textwidth\columnwidth\hsize\csname
@twocolumnfalse\endcsname
%--------------------------------------------------------------------
%--------------------------------------------------------------------

\title{Treating instabilities in a hyperbolic formulation of
Einstein's equations}
\author{Mark A.\ Scheel$^1$, Thomas W.\ Baumgarte$^2$,
        Gregory B.\ Cook$^1$,\\
        Stuart L.\ Shapiro$^{2,3}$, and Saul A.\ Teukolsky$^{1,4}$}
\address{$^1$Center for Radiophysics and Space
Research,Cornell University, Ithaca, New York\ \ 14853}
\address{$^2$Department of Physics, University of
Illinois at Urbana-Champaign, Urbana, IL\ \ 61801}
\address{$^3$Department of Astronomy and
NCSA, University of Illinois at Urbana-Champaign, Urbana, IL\ \ 61801}
\address{$^4$Departments of
Physics and Astronomy, Cornell University, Ithaca, New York\ \ 14853}
\date{\today}

\maketitle

\begin{abstract}
We have recently constructed a numerical code that evolves a
spherically symmetric spacetime using a hyperbolic formulation of
Einstein's equations. For the case of a Schwarzschild black hole, this
code works well at early times, but quickly becomes inaccurate on a
time scale of $10-100 M$, where $M$ is the mass of the hole.  We
present an analytic method that facilitates the detection of
instabilities. Using this method, we identify a term in the evolution
equations that leads to a rapidly-growing mode in the solution. After
eliminating this term from the evolution equations by means of
algebraic constraints, we can achieve free evolution for times
exceeding $10\,000M$.  We discuss the implications for
three-dimensional simulations.
%\narrowtext
\end{abstract}

%--------------------------------------------------------------------
%--------------------------------------------------------------------
% This is the other line to be uncommented for 2 column format
\vskip2pc]
%--------------------------------------------------------------------
%--------------------------------------------------------------------

\section{Introduction} \label{sec:intro}

When solving Einstein's equations as an initial value problem, one
considers spacetime as a foliation of spacelike hypersurfaces, or
``slices''.  Einstein's equations then separate into two types:
constraint equations, which relate the dynamical variables on each
particular slice, and evolution equations, which describe how these
variables propagate from one slice to the next.  The constraints are
analogous to the divergence equations in Maxwell's theory, and the
evolution equations are analogous to the curl equations.

As in Maxwell's theory, the evolution equations admit solutions that
violate the constraints.  However, if the constraints are satisfied on
the initial slice and on all spatial boundaries, then the evolution
equations guarantee that the constraints are satisfied elsewhere.
This permits numerical solution schemes in which only the evolution
equations are explicitly solved at each time step.

Such ``free evolution'' schemes are desirable for several reasons.
First, the constraints are typically nonlinear elliptic equations,
which are difficult and costly to solve on a computer, especially in
the general case of three spatial dimensions. Second, a free evolution
scheme allows one to track numerical errors by monitoring the
constraints at each time step.

For numerical evolution of black holes, an additional advantage of a
free evolution scheme is that one can, in principle, excise a black
hole from the spacetime and evolve only the exterior region, and one
can do so without imposing explicit boundary conditions on the
horizon. This is the basis for so-called ``apparent horizon boundary
condition'' methods, which are thought to be crucial for long-term
numerical evolution of black hole
spacetimes\cite{seidel_suen92,anninos_etal95,scheel95a,scheel95b,%
bona_masso95a,marsa96,scheel_etal97a,bbhprl98a,bbhprl98b}.  However,
excising a black hole from a spacetime is known to be mathematically
well-defined only if the evolution equations are hyperbolic and if the
characteristic curves of the hyperbolic system are ``physical'', that
is, if they lie within the local light cone. In this case, the
structure of the equations guarantees that no information, including
gauge information, can emerge from the hole. For non-hyperbolic
representations of general relativity such as the usual ADM\cite{ADM}
formulation, the evolution equations are of no mathematical type for
which well-posedness has been proven, so the suitability of these
formulations for black hole excision must be determined
empirically on a case-by-case basis.  It is in part for this reason
that much attention has been recently focused on hyperbolic
representations of Einstein's
equations\cite{frittelli_reula94,choquet_york95,abrahams_etal95,%
bona_masso95b,mvp96,frittelli_reula96,friedrich96,estabrook_etal97,%
anderson_etal98}.

A key stumbling block in numerical work, particularly in
finite-difference solutions of initial value problems, is the tendency
for numerical computations to become unstable. Instabilities have many
origins, and the cause of any particular instability found in a
numerical code is often difficult to deduce.  Furthermore, if the
desired analytic solution is unknown, it can be difficult to
distinguish between an instability and a case in which the analytic
solution simply grows without bound.  Examples of the latter include
systems that evolve to physical singularities ({\it e.g.,\/}
Oppenheimer-Snyder collapse evolved using geodesic slicing) and those
that evolve toward coordinate singularities ({\it e.g.,\/} a
Schwarzschild black hole evolved with maximal time slicing, and
several harmonic-slicing examples that become singular for certain
choices of the initial lapse
function\cite{alcubierre97,alcubierre_masso98}).  When diagnosing
instabilities in numerical simulations, it is therefore preferable to
study instances in which the analytic solution is known and
well-behaved.

We distinguish between two types of instabilities: a type in which the
numerical finite-difference equations admit rapidly-growing solutions
that do not satisfy the underlying continuum differential equations,
and a type in which the continuum equations themselves admit growing
modes that are absent in the desired solution but are excited by
numerical perturbations. An example of the former type, which we will
call a numerical instability because it depends on the numerical
finite-difference equations, is the well-known Courant instability
that can arise in explicit finite-difference solutions of hyperbolic
PDEs. The high-frequency modes that characterize a Courant instability
do not satisfy the underlying differential equations.

The latter type, which we will call a ``continuum'' instability
because the unstable mode satisfies the continuum differential
equations, commonly occurs in systems of equations that admit both
well-behaved and growing solutions. Although one might be interested
in the well-behaved solution, the growing mode eventually dominates if
it at any time acquires a nonzero amplitude via numerical errors.  A
simple example is the equation $\ddot y = y/9$ with initial conditions
$y = 1$ , $\dot y = -1/3$. For these initial conditions the unique
analytic solution is $y=e^{-t/3}$, but a naive numerical integration
of this problem is unstable as it proceeds forward in time because
numerical perturbations excite the growing solution $y=e^{t/3}$.

For numerical solutions of Einstein's equations, a continuum
instability may be due to a gauge mode excited by inaccuracies in
numerically-determined coordinate conditions.  Or, in the case of a
free evolution scheme, it may be caused by a rapidly-growing mode that
satisfies the evolution equations but violates the constraints.  This
latter case is possible despite the fact that the evolution equations
preserve the constraints, because in numerical computations neither
the evolution equations nor the constraints are {\it exactly\/}
satisfied. Constraint-violating modes have been discussed in the
literature\cite{frittelli97,gundlach_pullin97,burko_ori97} but their
importance for numerical free evolution schemes remains controversial.

Eliminating a continuum instability often requires a different
approach than removing a numerical one, because these two types of
instability stem from quite different sources.  To remove a numerical
instability, one must change the numerical algorithm (or details of the
algorithm such as the size of the time step) that is used to solve the
equations, so that this algorithm no longer introduces growing
modes. To remove a continuum instability, one must either remove the
numerical perturbations that excite the undesired solution of the
continuum equations, change the numerical scheme in order to damp out
this solution, or modify the continuum equations themselves (possibly
including the choice of gauge) so that no growing solution is present.

In this paper we examine instabilities in a numerical free evolution
code that solves a spherically symmetric black-hole spacetime.  Our
code, which has been described in detail
elsewhere\cite{scheel_etal97a}, is based on a hyperbolic formulation
of general relativity (the ``Einstein-Ricci'' or ``ER'' formulation)
originally proposed by Choquet-Bruhat and
York\cite{choquet_york95,abrahams_etal95}.  For short integration
times our code performs well, but we show in section~\ref{sec:unmod}
that for the case of a Schwarzschild black hole it becomes unstable
and terminates on a time scale of $10-100M$, where $M$ is the mass of
the hole.  This occurs even in a gauge in which the analytic solution
is regular at the horizon and time-independent.  The rate at which our
errors grow is independent of the numerical time discretization
$\Delta t$ and the spatial discretization $\Delta r$, suggesting that
the growth is due to a continuum instability rather than a numerical
one.

In section~\ref{sec:analysis} we present a method of analyzing the
evolution equations that facilitates the detection of continuum
instabilities.  In the simplest application of this method we consider
each ER evolution equation separately. For each equation, we examine
the free evolution of the ER variable governed by that equation,
treating all other ER variables as fixed and given by the
Schwarzschild solution.  We ask whether perturbations of the evolved
ER variable about its Schwarzschild value grow rapidly with time.  We
find that most of the ER equations, when treated individually in this
manner, are stable, but that one of the ER equations is sensitive to a
continuum instability.  A single term on the right-hand side of the
unstable equation is responsible for the growing mode.

In section~\ref{sec:mod} we construct a modified set of evolution
equations that no longer contain this troublesome term. This is done
primarily by using algebraic constraints to rewrite the right-hand
side of one equation.  We find that numerical free evolution of the
modified set of equations remains accurate for times in excess of
$10\,000M$. This substantial improvement indicates that the
rapidly-growing mode found by our analysis in
section~\ref{sec:analysis} is the dominant instability afflicting free
evolution of the unmodified ER equations.  In
section~\ref{sec:discussion} we discuss our method of stability
analysis and apply it to the three-dimensional Einstein-Ricci
equations, as well as to the Einstein-Bianchi\cite{anderson_etal98}
and ADM systems. We discuss the implications for three-dimensional
free evolution schemes.

\section{Equations}\label{sec:eqs}
\subsection{The ER Formalism} \label{sec:eqns.er}
Here we summarize the fundamental variables and equations used in the
ER representation of general relativity.  For details of the ER
formulation and a derivation of the equations, see
\cite{choquet_york95,abrahams_etal95}.

We write the metric in the usual 3+1 form
\begin{equation} \label{fourmetric}
ds^2=-N^2dt^2+g_{ij}(dx^i+\beta^idt)(dx^j+\beta^jdt),
\end{equation}
where $N$ is the lapse function, $\beta^i$ is the shift vector, and
$g_{ij}$ is the three-metric on a spatial hypersurface of constant
$t$.

Define the variables
\begin{mathletters}\label{er:variables}
\begin{eqnarray}
K_{ij} &\equiv& -{1\over 2}N^{-1}\dt g_{ij},\\
L_{ij} &\equiv& N^{-1}\dt K_{ij},\label{def:lij}\\
M_{kij}&\equiv& D_kK_{ij},\label{def:mkij}\\
a_i    &\equiv& D_i(\ln N),\label{def:ai}\\
a_{0i} &\equiv& N^{-1}\dt a_i,\\
a_{ij} &\equiv& D_ja_i\label{def:aij}.
\end{eqnarray}
\end{mathletters}
Here $D$ is the three-dimensional covariant derivative compatible with
the three-metric $g_{ij}$, the time derivative operator is
\begin{equation} \label{dt}
\dt\equiv {\partial\over\partial t}-\pounds_\beta,
\end{equation}
and \pounds\ denotes a Lie derivative. The quantity $K_{ij}$ is the
usual extrinsic curvature.  

The vacuum evolution equations for the general three-dimensional case
can be found in \cite{choquet_york95,abrahams_etal95,scheel_etal97a}.
The vacuum constraint equations include
\begin{mathletters} \label{er:cons}
\begin{eqnarray} 
0  &=& \bar{R}_{ij} - L_{ij} + H K_{ij} - 2K_{ik} K^k_j - a_i a_j - a_{ij},
         \label{er:cons:lij}\\
0  &=& L_i^i + K^{ij}K_{ij} + a^i a_i + a_i^i, \label{er:cons:ham}\\
0  &=& {M^j}_{ji} - {M_{ij}}^j, \label{er:cons:mom} \\
0  &=& a_{0i} + H a_i + {M_{ij}}^j, \label{er:cons:a0i}
\end{eqnarray}
where $\bar{R}_{ij}$ is the three-dim\-en\-sion\-al Ric\-ci tensor formed
from the three-metric $g_{ij}$. 
Eqs.~(\ref{er:cons:lij})--(\ref{er:cons:mom}) follow from the
Gauss-Codazzi-Ricci equations for embedding a foliation into a
higher-dimensional space, and Eq.~(\ref{er:cons:a0i}) follows from
harmonic time slicing.  Additional constraints that must be satisfied
at all times are the definitions~(\ref{def:mkij}), (\ref{def:ai}), and
(\ref{def:aij}), and the usual relation between $\Gamma^k{}_{ij}$ and
derivatives of $g_{ij}$.

\end{mathletters}

\subsection{Spherical Symmetry} \label{sec:eqns.ss}
The spherically symmetric three-metric can be written in the general form
\begin{equation} \label{3metric}
{}^{(3)}ds^2 = A^2dr^2+B^2r^2(d\theta^2+\sin^2\theta\,d\phi^2),
\end{equation}
where ($r,\theta,\phi$) are the usual spherical coordinates.  Define
\begin{mathletters} \label{erss:vars}
\begin{eqnarray}
\grt   &\equiv& 2 Br\Gamma^\theta{}_{\theta r}
              = 2 Br\Gamma^\phi{}_{\phi r}     
				   \nonumber\\
              &=&-{2A^2\over Br}\Gamma^r{}_{\theta\theta}
              =-{2A^2\over Br \sin^2\theta}\Gamma^r{}_{\phi\phi},\\
\at    &\equiv& a^\theta{}_\theta = a^\phi{}_\phi,\\
\lt    &\equiv& L^\theta{}_\theta = L^\phi{}_\phi,\\
\kt    &\equiv& K^\theta{}_\theta = K^\phi{}_\phi,\\
\mrt   &\equiv& M_r{}^\theta{}_\theta = M_r{}^\phi{}_\phi,\\
\mtr   &\equiv& M^\theta{}_\theta{}_r = M^\phi{}_\phi{}_r,
\end{eqnarray}
where the subscript $T$ denotes ``transverse''.  
\end{mathletters}

The evolution equations can be written in the form
\begin{mathletters}
\label{erss:ev}
\begin{eqnarray}
\dt A    &=& -NA\krtil,\label{erss:ev:a}\\
\dt Br   &=& -NBr\kt,\\
\dt \kr  &=& \phantom{-}N\lr,\\
\dt \kt  &=& \phantom{-}N(\lt+2\kt^2),\label{erss:ev:kt}\\
\dt N    &=& -N^2(\krtil+2\kt),\label{erss:ev:n}\\
\dt a_r  &=& \phantom{-}N\aor,\\
\dt \at  &=& \phantom{-}N\left[(2\mtr-\mrt-\ar\kt){\ar\over A^2}
                    \nonumber \right. \\ &&\left.
              +{\grt\over 2A^2Br}\aor
              +2\kt\at\right],\label{erss:ev:at}\\
\dt \gamr &=& -{N\over A^2}\left[\kr\ar+\mrr\right],\\
\dt \grt &=& -N\left[\kt\grt+2Br(\ar\kt+\mrt)\right],\\
\dt \mtr &=& \phantom{-}N\left[\kt(2\mtr+\mrt+\ar\kt)
                     \nonumber \right. \\ &&\left.
              +{\grt\over 2Br}\left({\lr\over A^2}-\lt\right)
                     \nonumber \right. \\ &&\left.
              +\krtil(2\mtr-\mrt-\ar\kt)\right],
              \label{erss:ev:mtr}
\\
\dt \arr   &=&   N\dr\aor
                     +    N\left[-\gamr\aor+\ar^2\krtil
				   \right.\nonumber \\ &&\left.
                           +{\ar\mrr\over A^2}+\ar\aor
			   \right],
                           \label{erss:ev:arr}\\
\dt \aor   &=&   {N \over A^2}\dr\arr
                     +    N\left[  {\mrr\over A^2}(2\kt-\krtil)
				   \right.\nonumber \\ &&\left.
		                   +\ar\left((\ar^2-\lr){1\over A^2}
				        -2\krtil(\krtil-3\kt)\right)   
				   \right.\nonumber \\ &&\left.
				   +{\arr\over A^2}\left(3\ar-2\gamr
						+{\grt\over Br}\right)
				   \right.\nonumber \\ &&\left.
				   +2\mrt\krtil
                           	   +\at\left(4\ar-{\grt\over Br}\right)
				\right],\\
\dt \mrr   &=&   N\dr\lr 
                     +    N\left[(\ar-2\gamr)\lr
				   \right.\nonumber \\ &&\left.
                           +2\krtil(\kr\ar+\mrr)\right],
				\label{erss:ev:mrr}\\
\dt \lr    &=&  {N \over A^2}\dr\mrr
                     +    N\left[\lr(4\kt-5\krtil)
				   +8\mrt\ar
				   \right.\nonumber \\ &&\left.
				   +{\mrr\over A^2}
				    \left(3(\ar-\gamr)+{\grt\over Br}\right)
                                   -{2\mtr\grt\over Br}
				   \right.\nonumber \\ &&\left.
				   +2\arr(3\kt-\krtil)
				   +\ar^2(10\kt-\krtil)
				   \right.\nonumber \\ &&\left.
				   -\kr(5\krtil^2-6\krtil\kt+2\kt^2)
				\right],\label{erss:ev:lrr}\\
\dt \mrt &=&  N\dr\lt
			\nonumber \\&&
                     +N\left[2\kt(\ar\kt+2\mrt)+\ar\lt\right],
	              \label{erss:ev:mrt}\\
\dt \lt &=&   {N\over A^2}\dr\mrt
        	         +N\left[\lt\krtil
	                 +{\ar^2\krtil\over A^2}
				   \right. \nonumber \\ &&
                         +(\lr+\arr)(\krtil-\kt){1\over A^2} 
			 -2\kt^3
				   \nonumber \\ &&
        	         +2\at(\krtil+\kt)
        	         -2\krtil^2\kt
			 +\krtil^3
				   \nonumber \\ &&
			 +2\krtil\kt^2
			 +{\mtr\over A^2}\left({\grt\over Br}-4\ar\right)
				   \nonumber \\ && \left.
			 +{\mrt\over A^2}
				\left({\grt\over Br}+3\ar-\gamr\right)
			  \right].\label{erss:ev:lt}
\end{eqnarray}
The constraints~(\ref{er:cons}) become
\end{mathletters}
\begin{mathletters}
\label{erss:cons}
\begin{eqnarray}
        {1\over Br}\left[-\dr\grt + \gamr\grt\right]
        -\arr-\ar^2
						\nonumber &&\\
	+\kr(2\kt-\krtil)-\lr &=& 0,
        \label{erss:cons:lrr}
     \\
      {1\over 2 A^2 Br}\left[-\dr\grt+\gamr\grt-{\grt^2\over 2 Br}\right]
						\nonumber &&\\
	     +{1\over B^2r^2}
	     +\kt\krtil-\at-\lt &=& 0,
        \label{erss:cons:lt}
     \\
     2\lt+{\lr\over A^2}+2\kt^2+\krtil^2+2\at
						\nonumber &&\\
                        +{1\over A^2}(\ar^2+\arr) &=& 0,
       \label{erss:cons:ham}
     \\ 
        \mrt - \mtr   &=& 0,
        \label{erss:cons:mom}
     \\	
        \aor + \ar(2\kt+\krtil)+{\mrr\over A^2}+2\mrt &=& 0.
	\label{erss:cons:a0r}
\end{eqnarray}
The additional constraints~(\ref{def:mkij}), (\ref{def:ai}),
(\ref{def:aij}), and the usual relation between $\Gamma^k{}_{ij}$ and
derivatives of $g_{ij}$ take the form
\begin{eqnarray}
        \dr\kr - 2\gamr\kr - \mrr &=& 0,
     \\
        \dr\kt - \mrt &=& 0,
     \\
        \mtr - {\grt\over 2Br}(\krtil - \kt) &=& 0,
        \label{erss:cons:mtr}
     \\
        \dr(\ln N) - \ar &=& 0,
     \\
        \at - {\grt\over 2A^2Br}\ar &=& 0,       
        \label{erss:cons:at}
     \\     
        \dr\ar - \arr - \gamr\ar &=& 0,
     \\
        \dr A - A\gamr &=& 0,
     \\
        \dr Br - {\grt\over 2} &=& 0.  
\end{eqnarray}
\end{mathletters}

\section{Free Evolution of ER system}
\label{sec:unmod}
\subsection{Method}\label{sec:unmod:method}
	We solve the spherically symmetric ER evolution
equations~(\ref{erss:ev}) at every time step using the causal
differencing method described in~\cite{scheel_etal97a}.  The
constraints are satisfied on the initial time slice but are not solved
explicitly during the evolution.

The inner boundary of the numerical domain is a surface that remains
within a grid spacing of the apparent horizon, $r=r_{AH}$. Because the
apparent horizon is an outgoing null or spacelike surface, the
hyperbolic evolution equations require no boundary condition there.
The outer boundary is an arbitrary spherical surface far from the
black hole at $r=r_{max}$.  At the outer boundary, we use the
``extended Robin'' condition discussed in~\cite{scheel_etal97a}.  This
outer boundary condition does not properly handle wavelike behavior,
but in practice it is adequate for the cases shown here.

The lapse function can be freely specified on the initial time slice,
and is subsequently determined by the harmonic time slicing condition
$\Box t = 0$.  The shift is chosen to satisfy the minimal strain
equation\cite{smarryork78}.  This equation minimizes the average
change in the three-metric as one evolves from one time slice to the
next, and is used to provide a shift vector that does not produce
coordinate singularities.  The minimal strain equation requires two
boundary conditions, for which we choose
\begin{eqnarray} 
\label{shiftbcinner}
\beta^r-{N\over A} &=& 0\qquad\hbox{at}\quad r=r_{AH},\\
\label{shiftbcouter}
\dr\left(r^2\beta^r\right) &=& 0\qquad\hbox{at}\quad r=r_{max}.
\end{eqnarray}
The inner boundary condition ensures that at the apparent horizon, the
coordinates move outward at the local speed of light, $c=N/A$. This
prevents the coordinates from falling into the black hole.  The outer
boundary condition ensures that the shift falls off like $r^{-2}$, in
accordance with the time-independent Schwarzschild solution written in
harmonic slicing (Eqs.~(\ref{shid}) below). We use a feedback
technique\cite{scheel_etal97a} to keep the horizon near $r=2M$.

\subsection{Initial data}\label{sec:unmod:initdata}
Our initial data are chosen on a time slice corresponding to a
well-behaved, fully time-independent harmonic foliation of the
Schwarzschild geometry
(cf. refs.\cite{bona_masso88,bona_masso89,cook_scheel97}).
Such a slice penetrates the event horizon without encountering a
coordinate singularity, and extends to the physical singularity at
$r=0$. With an appropriate choice of spatial coordinates on the slice,
all dynamical variables are time-independent\cite{cook_scheel97} and
are given by
\begin{mathletters}
\label{shid}
\begin{eqnarray}
\label{shid:a}
A^2           &=& \left(1+{2M\over r}\right)
                  \left[1+\left(2M\over r\right)^2\right],\\
B             &=& 1,\\
N             &=& {1\over A},\\
\beta^r       &=& {4N^2M^2\over r^2},\label{shid:beta}\\
\gamr         &=& -{N^2M\over r^2}
	              \left[1+{4M\over r} +{ 12M^2\over r^2}\right],\\
\grt          &=& 2,\\
\kt           &=&  {4NM^2\over r^3},\\
\kr           &=& -{4NM^2\over r^3}
		   \left[2+{3M\over r}
                          +{4M^2\over r^2}	
                          +{4M^3\over r^3}
		   \right],\\	
\mrr          &=& {4N^3M^2\over r^4}
		   \left[6+{18M\over r}
			  +{35M^2\over r^2}
			  +{40M^3\over r^3}
				   \right.\nonumber \\ &&\qquad\qquad\left.
			  +{56M^4\over r^4}
			  +{64M^5\over r^5}
			  +{48M^6\over r^6}
		   \right],\\
\mrt          &=& -{4N^3M^2\over r^4}
		   \left[3+{5 M\over r}
			  +{ 8M^2\over r^2}
			  +{12M^3\over r^3}
		   \right],\\
\mtr          &=& \mrt,\\
\lr           &=& -{16N^4M^4\over r^6}
		   \left[14+{42 M\over r} 
			   +{85 M^2\over r^2}
			   +{120M^3\over r^3}
				   \right.\nonumber \\ &&\qquad\qquad\left.
			   +{136M^4\over r^4}
			   +{128M^5\over r^5}
			   +{80 M^6\over r^6}
		   \right],\label{shid:lrr}\\
\lt           &=& {16N^4M^4\over r^6}
		   \left[1+{M\over r} 
			   -{4M^3\over r^3}
		   \right],\\
\ar           &=& {N^2M\over r^2}
	             \left[1+{4M\over r} +{ 12M^2\over r^2}\right],\\
\at           &=& {N^4M\over r^3}
	             \left[1+{4M\over r} +{ 12M^2\over r^2}\right],\\
\arr          &=& -{MN^4\over r^3}
		   \left[2+{13M\over r} 
			   +{56M^2\over r^2}
				   \right.\nonumber \\ &&\qquad\qquad\left.
			   +{40M^3\over r^3}
			   -{48M^5\over r^5}
		   \right],\\
\aor          &=& {16M^3N^5\over r^5}
		  \left[1+{6M\over r} 
			   +{24M^2\over r^2}
				   \right.\nonumber \\ &&\qquad\qquad\left.
			   +{24M^3\over r^3}
			   +{16M^4\over r^4}
		   \right],
\end{eqnarray}
where $M$ is the mass of the black hole.  One can explicitly check the
time-independence of this solution by inserting~(\ref{shid}) into the
ER evolution equations~(\ref{erss:ev}) and verifying that all time
derivatives are zero. Note that~(\ref{shid}) satisfies the minimal
strain shift condition, as does any time-independent solution of
Einstein's equations.
\end{mathletters}

\subsection{Results}\label{sec:unmod:results}

\begin{figure}[!htb]
\begin{center}
\begin{picture}(240,250)
\put(0,0){\epsfxsize=3.5in\epsffile{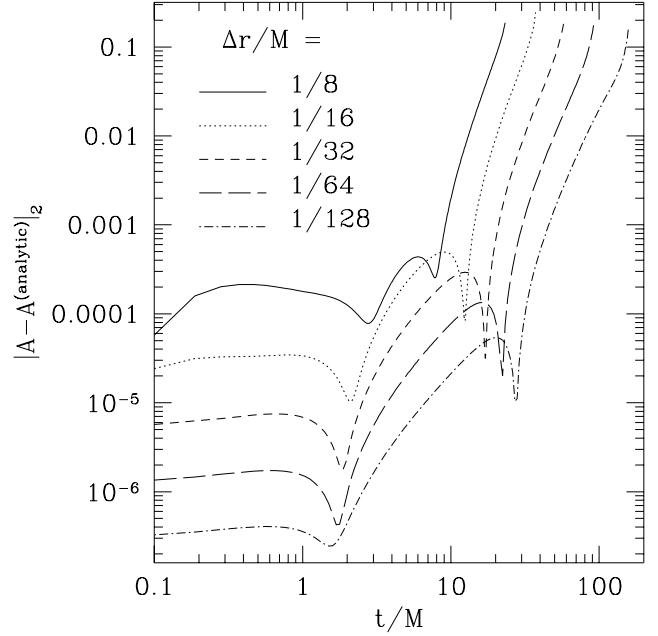}}
\end{picture}
\end{center}
\caption{The $\ell_2$ norm of $A$ minus its analytic
solution~(\ref{shid:a}), shown as a function of time for five grid
resolutions. The outer boundary is at $r_{max}=64M$ and the Courant
factor $\Delta t/\Delta r$ is $3/4$.  All five plots
terminate when the code crashes.
\label{fig:unmod:a}
}
\end{figure}

Figure~\ref{fig:unmod:a} shows the error in the metric function $A$ as
a function of time.  We plot the quantity $|A-A^{\rm an}|_2$, where
$A^{\rm an}$ is the analytic value of $A$ given by~(\ref{shid:a}), and
the $\ell_2$ norm of a quantity $q$ is defined by
\begin{equation}
|q|_2\equiv\sqrt{\sum_{i=1}^N q_i^2\over N},
\end{equation} 
The sum is over all grid points that contain valid data ({\it i.e.,\/}
all grid points outside the horizon).  The quantity $|A-A^{\rm an}|_2$
is shown for several different grid resolutions, each with the same
Courant factor $\Delta t/\Delta r$.

At early times, the error in $A$ varies with resolution like
${\mathcal O}(\Delta r)^2$, as expected for our second-order
convergent numerical method.  However, after about $10$--$30M$ the
error grows rapidly, approximately like $t^4$ at late times.  The
growth rate is independent of the grid resolution. Eventually, when
errors have become sufficiently large, the code crashes, typically
because it fails to locate an apparent horizon.

It is common for numerical finite-difference schemes to produce
solutions with errors that grow as truncation error
accumulates. However, such growth is typically linear in time, with a
slope proportional to $(\Delta t)^2$ (for a second-order scheme), and
can be easily defeated by increasing the resolution.  In contrast,
Figure~\ref{fig:unmod:a} shows a more rapid growth rate that increases
with time, indicating that we are observing something other than
accumulating truncation errors.

\begin{figure}[!htb]
\begin{center}
\begin{picture}(240,250)
\put(0,0){\epsfxsize=3.5in\epsffile{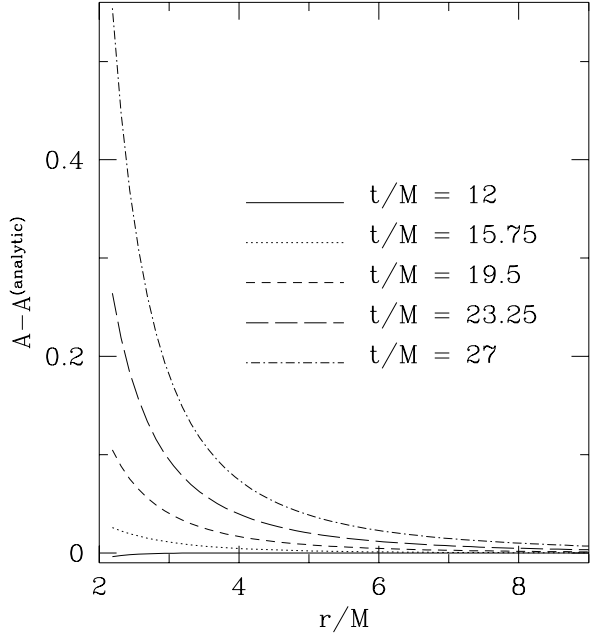}}
\end{picture}
\end{center}
\caption{Error in $A$ as a function of coordinate radius, for the
$\Delta r/M = 1/16$ case shown in Figure~\ref{fig:unmod:a}. The
function $A-A^{\rm an}$ is plotted at five times.  The error grows
rapidly but smoothly until the code crashes.
\label{fig:unmod:avsr}
}
\end{figure}

In Figure~\ref{fig:unmod:avsr} we plot the error in $A$ as a function
of radius for several different times. The error is greatest near the
horizon and remains smooth in both space and time as it grows. The
fact that our errors are largest near the black hole does not
necessarily indicate that the instability is somehow associated with
our treatment of the inner boundary; one expects numerical errors to
be greater for smaller values of $r$ simply because most quantities
in~(\ref{shid}) behave like $1/r^n$ with positive $n$.

\begin{figure}[!htb]
\begin{center}
\begin{picture}(240,250)
\put(0,0){\epsfxsize=3.5in\epsffile{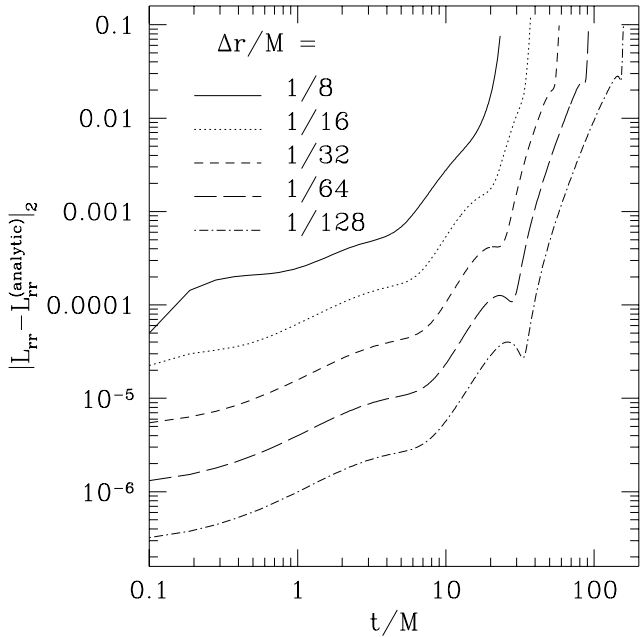}}
\end{picture}
\end{center}
\caption{The $\ell_2$ norm of $L_{rr}$ minus its analytic
solution~(\ref{shid:lrr}) as a function of time, shown for the same
cases as plotted in Figure~\ref{fig:unmod:a}.
\label{fig:unmod:lrr}
}
\end{figure}

\begin{figure}[!htb]
\begin{center}
\begin{picture}(240,250)
\put(0,0){\epsfxsize=3.5in\epsffile{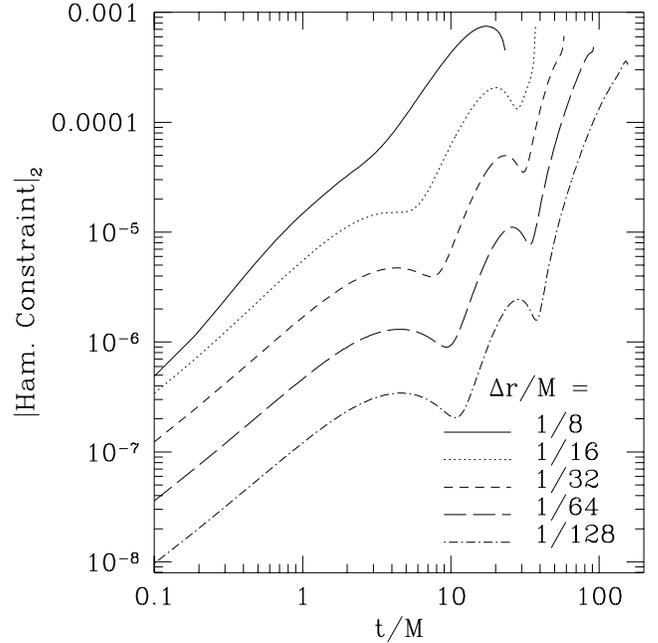}}
\end{picture}
\end{center}
\caption{The $\ell_2$ norm of the Hamiltonian
constraint~(\ref{erss:cons:ham}) versus time, shown for the same five
cases as in Figure~\ref{fig:unmod:a}.
\label{fig:unmod:ham}
}
\end{figure}

Other quantities behave much like the error in $A$.  In
Figure~\ref{fig:unmod:lrr} we plot the error in $\lr$ with respect to
the analytic solution~(\ref{shid:lrr}), and in
Figure~\ref{fig:unmod:ham} we plot the left-hand side of the
Hamiltonian constraint~(\ref{erss:cons:ham}).  Both quantities
are approximately second-order convergent, but at late times they
increase rapidly (faster than linearly) in time at a rate independent
of the grid resolution.

\begin{figure}[!htb]
\begin{center}
\begin{picture}(240,250)
\put(0,0){\epsfxsize=3.5in\epsffile{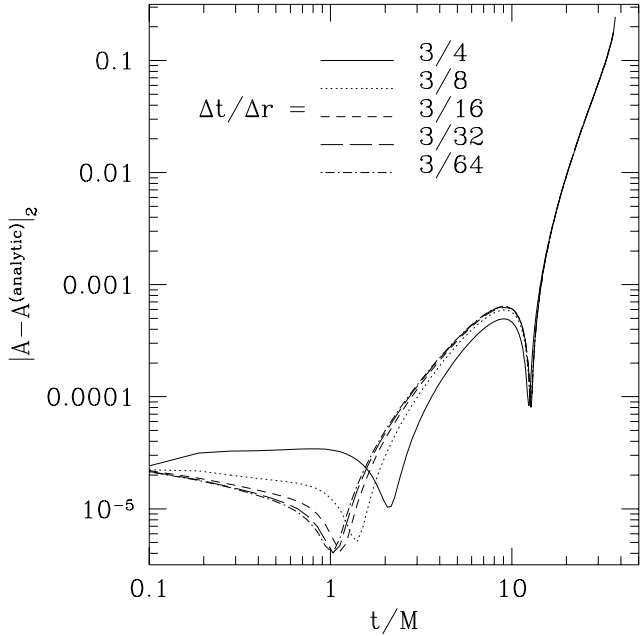}}
\end{picture}
\end{center}
\caption{The $\ell_2$ norm of $A-A^{\rm an}$ versus time shown for
five different values of $\Delta t$, each with $\Delta r/M =
1/16$. The outer boundary is at $r=64M$.  At late times, the dominant
error is independent of $\Delta t$.
\label{fig:unmod:adt}
}
\end{figure}

Figures~\ref{fig:unmod:a}--\ref{fig:unmod:ham} suggest that the
instability is not purely numerical.  Numerical instabilities
typically grow like $e^{n}$, where $n$ is the number of time
steps. Consequently, for a numerical instability one expects that
reducing the time discretization $\Delta t$ would make the instability
grow {\em faster} as a function of time, because integrating to a
particular value of $t$ requires more steps.  However, in
Figures~\ref{fig:unmod:a}--\ref{fig:unmod:ham}, $\Delta t$ is
decreased with each finer grid resolution, but the growth rate is
unaffected. Similarly, at late times we see no change in the growth
rate if we vary $\Delta t$ while keeping the grid resolution fixed, as
shown in Figure~\ref{fig:unmod:adt}. Instead, for $\Delta t\to 0$ our
errors converge to a limit (this is simply the limit in which
numerical truncation error is dominated by $\Delta r$ instead of
$\Delta t$).

Our results instead suggest that our code suffers from a continuum
instability. In this case, the code should remain second-order
convergent and the growth rate of errors should depend only on the
continuum equations and not on numerical parameters like $\Delta r$ or
$\Delta t$.  A smaller $\Delta t$ or $\Delta r$ should not intensify
the instability, but instead should improve our simulations by virtue
of reducing the numerical perturbations that excite the offending
mode. Our results are consistent with these expectations.

One possible source of a continuum instability is a rapidly-increasing
constraint-violating solution of the evolution equations that is being
excited by numerical perturbations.  Another is a gauge mode that is
not present in the analytic solution.  In the case of a gauge mode,
one would expect gauge-invariant quantities to remain relatively
unaffected while other quantities blow up. However, at late times, both
gauge-dependent quantities like $\lr$ (Figure~\ref{fig:unmod:lrr}) and
gauge-invariant quantities like the Hamiltonian constraint
(Figure~\ref{fig:unmod:ham}) increase rapidly with time at
approximately the same rate.

\section{Stability of individual evolution equations}
\label{sec:analysis}
To gain further insight into the nature of the instability, we
consider each ER evolution equation separately. For each evolution
equation, we treat the ER variable governed by that equation as
freely evolving, but we fix the remaining ER variables to the analytic
expressions given in~(\ref{shid}).  In this way we can study the
stability of each individual evolution equation in the absence of all
couplings to other equations.  Although this analysis will not shed
light on any instabilities that are caused by these couplings, it is
likely that if any of the evolution equations are found to be unstable
individually, they will remain unstable when coupled to the other
equations.  

We note that the method of analysis described below can also be used
to examine coupled sets of equations as long as the couplings do not
arise from derivative terms---this is described in more detail in
Appendix~\ref{sec:append:a}. However, we will see that treating one
equation at a time is sufficient for the case discussed here.

Let $y$ represent any of the ER variables that evolve according
to~(\ref{erss:ev}). If all ER variables other than $y$ are considered
known functions of $r$, then the evolution equation for $y$ takes the form
\begin{equation}
{\partial\over\partial t} y - \beta(r) \dr y = S(y,r),\label{advective}
\end{equation}
where the function $S(y,r)$ contains no derivatives of $y$.
If we perturb the quantity $y$ about its
time-independent solution by writing $y \to y + \xi$,
then~(\ref{advective}) yields, to first order in $\xi$,
\begin{equation}
{\partial\over\partial t} \xi - \beta(r)\dr\xi = R(r)\xi,\label{padvective}
\end{equation}
where $R(r)$ does not depend on $\xi$.  

For each of the ER evolution equations~(\ref{erss:ev}) there is a
corresponding perturbation equation of the form~(\ref{padvective}).
Each perturbation equation has a different function $R(r)$ that
depends on the right-hand side of the corresponding evolution
equation.  We will see that the form of $R(r)$ is what determines
whether a particular evolution equation is individually stable.

For the simple case in which $\beta(r)$ and $R(r)$ are constants and
$\beta>0$, the solution to~(\ref{padvective}) on $r\in[2M,\infty]$ is
\begin{equation}
\xi(r,t)=\xi_0(r+\beta t)e^{Rt},\label{padvectivesoln}
\end{equation}
where $\xi_0(r)$ is the initial perturbation at $t=0$.  The stability
is determined by the sign of $R$: If $R>0$ (assuming that the initial
perturbation falls off with radius more slowly than $e^{-rR/\beta}$),
the perturbation grows exponentially with time; if $R<0$ (assuming
that the initial perturbation grows with radius more slowly than
$e^{r|R|/\beta}$), the perturbation decays.

For the more realistic case of nonconstant $R$ and $\beta$, the
solution to~(\ref{padvective}) is more complicated
than~(\ref{padvectivesoln}) and is considered in
Appendix~\ref{sec:append:a}. Nevertheless, one can roughly determine
whether a given ER evolution equation is individually stable by
examining the sign of the function $R(r)$ associated with that
evolution equation.

Applying this criterion to the ER evolution equations~(\ref{erss:ev}),
we find that $R(r)$ is everywhere negative for all but
four of these equations, indicating that these equations
should be stable to small perturbations.
The four remaining equations
have positive $R(r)$, suggesting that they might be unstable.
If $R(r)_{[y]}$ denotes the function $R(r)$ associated with
perturbations of the variable $y$, then the four positive $R(r)_{[y]}$
are
\begin{mathletters}
\label{fourrhs}
\begin{eqnarray}
R(r)_{[K_T]}    &=& 4N\kt = {2z^3\over M(1+z)(1+z^2)},
		\label{fourrhs:kt}\\
R(r)_{[a_T]}    &=& 2N\kt = { z^3\over M(1+z)(1+z^2)},
		\label{fourrhs:at}\\
R(r)_{[M_{rT}]} &=& 4N\kt+\dr\beta 
		\nonumber\\ &=&
		{z^3(2+3z+4z^2+5z^3)\over 2M(1+z)^2(1+z^2)^2},
		\label{fourrhs:mrt}\\
R(r)_{[L_{rr}]} &=& N(4\kt-5\krtil)+2\dr\beta 
		\nonumber\\ &=&
		{z^3(20+19z+18z^2+17z^3)\over 4M(1+z)^2(1+z^2)^2},
		\label{fourrhs:lrr}
\end{eqnarray}
where $z\equiv 2M/r$ and the expressions in terms of $z$ have been
obtained from the analytic solution~(\ref{shid}).
\end{mathletters}

We can test whether perturbations of individual evolution equations
are indeed unstable by modifying our code so that a single dynamical
variable may be evolved in time while all other variables, including
the shift, are held fixed to the analytic solution~(\ref{shid}).  We
find numerically that all evolution equations~(\ref{erss:ev}) are
individually stable except~(\ref{erss:ev:lrr}), the equation for
$\lr$.

Our above analysis predicted that the $\lr$ equation should be
individually unstable because it is associated with a positive
$R(r)$. However, it also predicted that the $\kt$, $\at$, and $\mrt$
equations should be unstable for the same reason. As shown by a more
detailed analysis in Appendix~\ref{sec:append:a}, the $\kt$, $\at$,
and $\mrt$ equations are stable because their corresponding values of
$R(r)$ are much smaller in magnitude than the value of $R(r)$
associated with the $\lr$ equation.

The growing mode allowed by the $\lr$ evolution
equation~(\ref{erss:ev:lrr}) can be described as a continuum
instability: it depends only on the equation itself and the
equilibrium solution, and not on numerics. The only role of numerics
is to produce the initial perturbations that excite the unstable mode.

\section{Modified evolution equations}\label{sec:mod}
\subsection{Modifications for stability}\label{sec:mod:method}

The large positive $R(r)$ associated with perturbations of $\lr$
originates from the term $N\lr(4\kt-5\krtil)$ that appears on the
right-hand side of the $\lr$ evolution equation~(\ref{erss:ev:lrr}).
This term must be modified if the $\lr$ evolution equation is to be
made individually stable.  There are several ways to accomplish this.

One possibility is to change variables.  If one evolves some quantity
$Q\lr$ instead of $\lr$, where $Q$ is some combination of the other ER
variables, then perturbations of $Q\lr$ will be governed
by~(\ref{padvective}) with some new value of $R(r)$. By careful choice of
$Q$ one hopes to obtain a more favorable (more negative) $R(r)$.  For
example, the evolution equation for the quantity $B^2r^2\lr$ yields
$R(r)=N(2\kt-5\krtil)+2\partial\beta/\partial r$, which is still
positive but is slightly smaller in magnitude
than~(\ref{fourrhs:lrr}). Similarly, the evolution equation for
$L^r_r$ yields $R(r)=N(2\kt-3\krtil)$. However, there are two reasons
why such a procedure is unattractive as the sole method of stabilizing
the $\lr$ equation. First, the ER equations are linear in $L_{ij}$,
$M^k_{ij}$, $a_{ij}$, and $a_{0i}$ (but nonlinear in the other
variables), and evolving $Q\lr$ where $Q$ is anything other than the
metric functions or the lapse would spoil this linearity.  Second, in
order to make $R(r)$ nonpositive everywhere by evolving the quantity
$B^nr^n\lr/A^m$, it turns out that the required value of $n$ is large
enough that $B^nr^n\lr/A^m$ grows with $r$, hampering our ability to
impose an accurate outer boundary condition.

Another approach is to use the constraint equations to eliminate the
troublesome term that appears on the right-hand side of the $\lr$
evolution equation~(\ref{erss:ev:lrr}).  In order to avoid changing
the hyperbolic character of the evolution equations, one must use only
constraint equations that are algebraic, that is, those that contain
no derivatives. Fortunately, many of the ER constraints are
algebraic. For some constraints this is merely a result of spherical
symmetry, but several ER constraint equations are algebraic even in
the general case of three spatial dimensions plus time.  In spherical
symmetry, the algebraic constraints are Eqs.~(\ref{erss:cons:ham}),
(\ref{erss:cons:mom}), (\ref{erss:cons:a0r}), (\ref{erss:cons:mtr}),
and~(\ref{erss:cons:at}).  An additional algebraic constraint
can be formed from~(\ref{erss:cons:lrr}) and~(\ref{erss:cons:lt}) by
eliminating the derivative of $\grt$, yielding.
\begin{eqnarray}
\label{erss:cons:nogrt}
     2\lt-{\lr\over A^2}-\krtil^2+2\at-{1\over A^2}(\ar^2+\arr) 
		\nonumber&&\\
	 -{2\over B^2r^2}+{\grt^2\over 2 A^2 B^2 r^2} &=& 0.
\end{eqnarray}
Because we wish to modify the $\lr$ term on the right-hand side
of~(\ref{erss:ev:lrr}) for the case in which all variables except
$\lr$ are fixed to the analytic solution, the only relevant algebraic
constraints are those that involve $\lr$, namely~(\ref{erss:cons:ham})
and~(\ref{erss:cons:nogrt}).

We have found several methods of obtaining an individually stable
evolution of $\lr$. These all involve the use of algebraic constraint
equations, and some also employ a change of variables.  We have had
the most success with the following approach: First eliminate $\lt$
from~(\ref{erss:cons:ham}) and~(\ref{erss:cons:nogrt}) to obtain
\begin{equation}
\label{erss:cons:lrss}
{\grt^2-4A^2\over 4 B^2 r^2}-\ar^2-\arr-A^2(\kt^2+\krtil^2)-\lr = 0.
\end{equation}
Then write down the evolution equation for the quantity $L^r_r$,
and add $N(4\kt-5\krtil)/A^2$ times Eq.~(\ref{erss:cons:lrss}) to the
right-hand side, yielding
\begin{mathletters}
\label{erss:ev:ups}
\begin{eqnarray}
\dt L^r_r     &=&  {N \over A^2}\dr M^r_{rr} +N\left[ 
		   2\krtil L^r_r
 		   +8{\mrt\over A^2}\ar
			\right.\nonumber\\&&\left.
		   +M^r_{rr}\left(3\ar-\gamr+{\grt\over Br}\right)           
		   -2{\mtr\grt\over A^2Br}
			\right.\nonumber\\&&\left.
	           +(5\krtil-4\kt){1\over B^2r^2}
			\left(1-{\grt^2\over 4A^2}\right)
			\right.\nonumber\\&&\left.
		   +{\ar^2\over A^2}(6\kt+4\krtil)
		   +{\arr\over A^2}(2\kt+3\krtil)
			\right.\nonumber\\&&\left.
		   +\kt(2\krtil^2+3\kt\krtil-4\kt^2)
  		   \right].\label{erss:ev:lup}
\end{eqnarray}
Because we now evolve $L^r_r$ instead of $\lr$, we also choose to
evolve $M^r_{rr}$ instead of $\mrr$. This preserves the symmetry
between the $L$--$M$ pairs of evolution equations that make up wave
equations.  The evolution equation for $M^r_{rr}$ is
\begin{eqnarray}
\dt M^r_{rr}  &=&   N\dr L^r_r 
			\nonumber\\&&
                     +N\left[\ar L^r_r+2\krtil^2\ar+4\krtil M^r_{rr}
		     \right].\label{erss:ev:mup}
\end{eqnarray}
Evolving $M^r_{rr}$ has an additional advantage: perturbations of
$M^r_{rr}$ governed by~(\ref{erss:ev:mup}) have a smaller (more
negative) $R(r)$ then perturbations of $\mrr$ governed
by~(\ref{erss:ev:mrr}), so perturbations of $M^r_{rr}$ should decay
more rapidly.
\end{mathletters}

\subsection{Results}\label{sec:mod:results}

\begin{figure}[htb]
\begin{center}
\begin{picture}(240,250)
\put(0,0){\epsfxsize=3.5in\epsffile{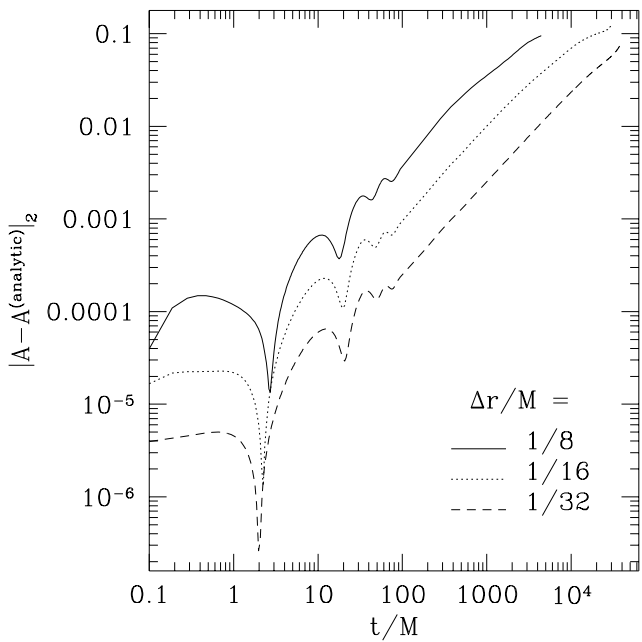}}
\end{picture}
\end{center}
\caption{The $\ell_2$ norm of the error in $A$ versus time, computed
for three resolutions using the modified evolution equations.  The
outer boundary is at $r_{\rm max}=128M$ and $\Delta t/\Delta r = 3/4$.
For $t\protect\gtrsim 5M$ the growth is only linear in time, and the
code runs much longer than for the case shown in
Figure~\ref{fig:unmod:a}.
\label{fig:mod:a}
}
\end{figure}

\begin{figure}[htb]
\begin{center}
\begin{picture}(240,250)
\put(0,0){\epsfxsize=3.5in\epsffile{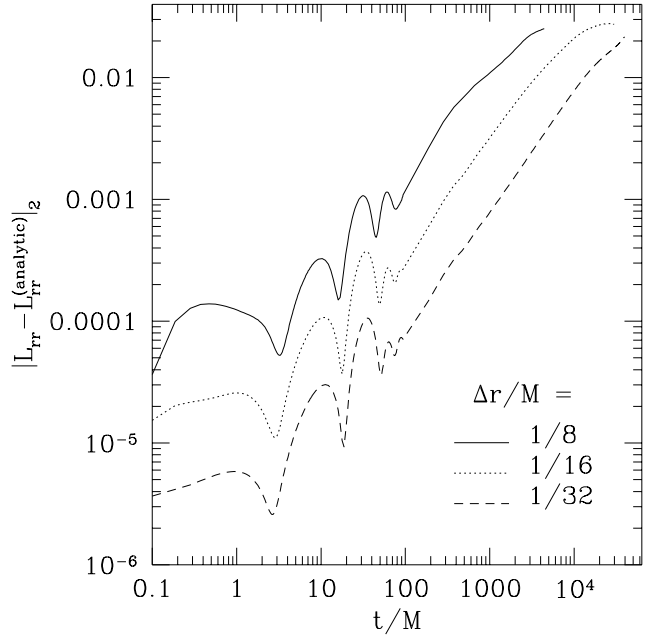}}
\end{picture}
\end{center}
\caption{The $\ell_2$ norm of the error in $\lr$ versus time, shown
for the same three cases as in Figure~\ref{fig:mod:a}.
\label{fig:mod:lrr}
}
\end{figure}

\begin{figure}[htb]
\begin{center}
\begin{picture}(240,250)
\put(0,0){\epsfxsize=3.5in\epsffile{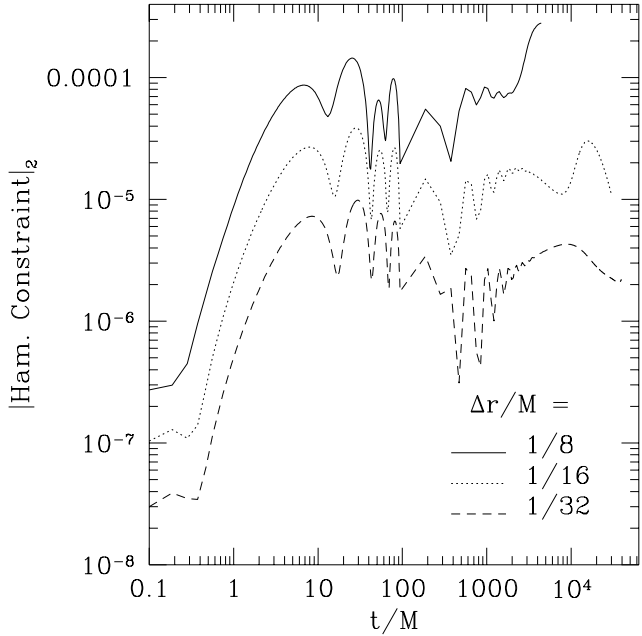}}
\end{picture}
\end{center}
\caption{The $\ell_2$ norm of the Hamiltonian constraint versus time, shown
for the same three cases as in Figure~\ref{fig:mod:a}. There
is no significant growth at late times.
\label{fig:mod:ham}
}
\end{figure}

Figures~\ref{fig:mod:a}--\ref{fig:mod:ham} show the $\ell_2$ norms of
the error in $A$, the error in $\lr$, and the Hamiltonian constraint
for simulations in which we solve the modified evolution
equations~(\ref{erss:ev:ups}) in place of~(\ref{erss:ev:mrr})
and~(\ref{erss:ev:lrr}).  The numerical method used in these
simulations is identical to the one used to integrate the unmodified
evolution equations in section~\ref{sec:unmod}. We use a larger outer
boundary radius, $r_{\rm max}=128M$, to suppress outer boundary
difficulties that become important at late times.

For the same grid resolution, our code integrates several orders of
magnitude farther in time when using the modified evolution equations
than when using the unmodified ones.  The large errors that grow on a
time scale of $10-100M$ in
Figures~\ref{fig:unmod:a}--\ref{fig:unmod:adt} are not present in
Figures~\ref{fig:mod:a}--\ref{fig:mod:ham}. Instead, numerical errors
increase linearly with time (or slower than linearly) for over
$10\,000M$ until difficulties associated with our treatment of the outer
boundary eventually halt the simulation.  

\begin{figure}[htb]
\begin{center}
\begin{picture}(240,250)
\put(0,0){\epsfxsize=3.5in\epsffile{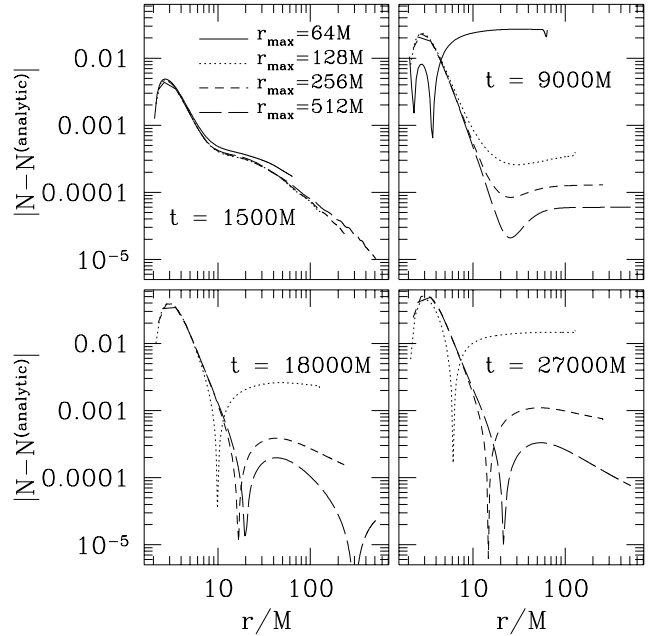}}
\end{picture}
\end{center}
\caption{The absolute value of the error in $N$ as a function of
radius, shown at various times for several cases of differing $r_{\rm
max}$. All plots have $\Delta r/M = 1/32$ and $\Delta t/\Delta r =
3/4$. The simulation with $r_{\rm max}=64M$ crashes at $12\,000M$.
\label{fig:mod:Nr}
}
\end{figure}

The errors in all dynamical variables except $N$ and $Br$ exhibit the
same linear growth as seen in Figures~\ref{fig:mod:a}
and~\ref{fig:mod:lrr}. Errors in $N$ and $Br$ are instead dominated by
outer-boundary effects that grow rapidly and eventually terminate our
code. Figure~\ref{fig:mod:Nr} shows the error in the lapse function
$N$ at various times, plotted as a function of radius for several
simulations with different outer boundary radii $r_{\rm max}$ but with
the same grid resolution $\Delta r$ and time discretization $\Delta
t$.  Increasing the outer boundary radius suppresses the rapid growth
of outer-boundary-related errors at late times and allows for much
longer simulations. It should also be possible to improve our results
by modifying our outer boundary condition, but the integration times
achieved by our code are already beyond what should be necessary for
modeling interesting 3D astrophysical problems such as black-hole
binary coalescence.

\section{Discussion}\label{sec:discussion}

The success of our free evolution scheme when solving the modified ER
equations is strong evidence that the growing continuum mode
identified in section~\ref{sec:analysis} is responsible for the
instability discussed in section~\ref{sec:unmod:results}.  The key
modification required to suppress the instability was the removal of a
term on the right-hand side of the $\lr$ equation, the very term that
our analysis in section~\ref{sec:analysis} singled out as problematic.
Although we have also improved the performance of our code by using
$L^r_r$ and $M^r_{rr}$ as dynamical variables instead of $\lr$ and
$\mrr$, we have verified that making this change of variables alone,
without removing the troublesome term in the $L^r_r$ equation by means
of constraints, yields results only marginally better than those shown
in section~\ref{sec:unmod:results}. Conversely, removing the unstable
term and evolving $\lr$ and $\mrr$ instead of $L^r_r$ and $M^r_{rr}$
still allows evolutions to thousands of $M$.

It is no surprise that the detailed behavior of the instability shown
in section~\ref{sec:unmod:results} is much more complicated than that
predicted by our simple analysis in section~\ref{sec:analysis} and in
Appendix~\ref{sec:append:a}.  We considered the evolution of a single
variable according to a single linear advective equation that
possesses only fixed, ingoing (for $\beta>0$) characteristic
curves. The ER system is actually a coupled system of nonlinear
advective and wave equations, and its three families of characteristic
curves (along the ingoing and outgoing light cone, and along the
normal to the foliation of time slices) depend on the solution. One
could do better than our treatment in section~\ref{sec:analysis} by
linearizing~(\ref{erss:ev}) about the analytic solution and
solving the entire system of coupled linear partial differential
equations; however, our approach is far simpler and appears to give
the correct qualitative results.

We emphasize that the results presented in
section~\ref{sec:mod:results} were obtained using free evolution, and
that no constraints have been enforced. Furthermore, we note that the
modifications discussed in section~\ref{sec:mod} do not alter the
hyperbolic character of the system. A different version of our code
evolves~(\ref{erss:ev}) while enforcing several algebraic constraint
equations (specifically, we solve~(\ref{erss:cons:mom}) for $\mrt$,
(\ref{erss:cons:nogrt}) and~(\ref{erss:cons:ham}) for $\lt$,
and~(\ref{erss:cons:a0r}) for $\aor$ after every time step), and
yields evolutions accurate for times on the order of $1000M$.  While
constraint enforcement allows our simulations to remain accurate for
far longer times than with free evolution of the unmodified ER
equations~(\ref{erss:ev}), our partially-constrained method eventually
succumbs to an instability slightly after $1000M$. The details of
exactly how constraint enforcement suppresses the continuum instability
found in section~\ref{sec:analysis} are unknown.

We have concentrated on a case in which the analytic solution is
manifestly time-independent, namely, when initial data given by
Eqs.~(\ref{shid}) are evolved using a harmonic time coordinate.
However, modifying our evolution equations also dramatically improves
our numerical results when initial data are chosen on a
minimally-modified ingoing Eddington-Finkelstein (MMIEF)\cite{marsa96}
time slice, so that subsequent evolution using harmonic time slicing
yields a time-dependent result.  Using our partially-constrained code,
we have shown\cite{cook_scheel97} that the evolution of MMIEF initial
data using harmonic time slicing relaxes to the solution~(\ref{shid})
at late times. The same result holds for free evolution
of the modified ER equations.

It is straightforward to extend the analysis in
section~\ref{sec:analysis} to the three-dimensional ER system. In this
case, it is useful to include couplings between tensor components. For
example, three-dimensional perturbations of $L_{ij}$, with all other
quantities held constant, obey
\begin{equation}
{\partial\over\partial t}\xi - \beta^i{\partial\over\partial x^i}\xi 
            = R(x,y,z)\xi,\label{padvective3d}
\end{equation}
which is similar to~(\ref{padvective}) except that here $\xi$ is a
column vector containing $(\delta L_{xx},\delta L_{xy},\ldots,\delta
L_{zz})$ and $R(x,y,z)$ is a matrix.  For perturbations about the
spherically-symmetric solution~(\ref{shid}), we find that the largest
eigenvalue of $R(x,y,z)$ is given by the same
expression~(\ref{fourrhs:lrr}) as in the spherically symmetric case,
indicating that the three-dimensional ER equations should suffer from
the same instability as their spherically-symmetric counterparts.
Applying the same analysis to the $K_{ij}$ evolution equation in the
ADM system (using the same analytic solution~(\ref{shid})) results in
eigenvalues of $R(x,y,z)$ that are the same size as
Eq.~(\ref{fourrhs:at}) and applying it to the $E_{ij}$, $H_{ki,j}$,
$K_{ij}$, and $\Gamma^k_{ij}$ equations in the Einstein-Bianchi
system\cite{anderson_etal98} yields eigenvalues of $R(x,y,z)$ that are
no larger than $3/2$ the size of Eq.~(\ref{fourrhs:at}), so we expect
that the type of continuum instability we find in the ER system
should not be present in either of these two other formalisms.

Although our stability analysis makes use of the analytic
solution~(\ref{shid}), in principle any other solution can be used
instead as a background for perturbations. Because the form of the ER
evolution equations given by~(\ref{erss:ev}) assumes harmonic slicing,
the only relevant time-independent solution is~(\ref{shid}). However,
for the case of the Einstein-Bianchi or ADM system evolved using a
different gauge, one might be interested in a different background
solution. The features of the background solution that are important
for determining stability are the signs and relative magnitudes of
components of $K_{ij}$ and derivatives of $\beta^i$. We note that
these features are approximately the same for the Schwarzschild
solution on time-independent MMIEF slices as they are for the
Schwarzschild solution on time-independent harmonic slices, so one
obtains similar stability criteria in both cases.

In the case of the ER equations, we are fortunate to have algebraic
constraints that can be used to modify the evolution equations without
affecting the hyperbolic character of the system, even in three
dimensions. However, not all the ER constraints are algebraic, and it
is unclear in the three-dimensional case which constraints must be
used in order to suppress instabilities.  In particular,
Eq.~(\ref{erss:cons:lrss}), which seems necessary for removing the
growing mode, is not algebraic in three dimensions.  This is because
Eq.~(\ref{erss:cons:lrss}) results from eliminating second derivatives
of the metric from Eqs.~(\ref{erss:cons:lrr})
and~(\ref{erss:cons:lt}); the three-dimensional equivalent is forming
a linear combination of components of Eq.~(\ref{er:cons:lij}) that
eliminates all second derivatives of $g_{ij}$ appearing in the Ricci
tensor $\bar{R}_{ij}$, and is not possible for a general spacetime.

One might ask why we do not use~(\ref{erss:cons:ham}) instead
of~(\ref{erss:cons:lrss}) to obtain a stable evolution equation for
$L^r_r$, since~(\ref{erss:cons:ham}) is algebraic in the general
three-dimensional case.  The answer is that it {\it is\/} possible to
use~(\ref{erss:cons:ham}) to obtain an {\it individually\/} stable
evolution equation for $L^r_r$.  However, doing so introduces a term
containing $\lt$ on the right-hand side of the $L^r_r$ evolution
equation, where no such term existed previously. This term generates a
continuum instability in the {\it coupled\/} $L^r_r$--$\lt$ system
(where all variables except $L^r_r$ and $\lt$ are held fixed to the
analytic solution).

To better understand why~(\ref{erss:cons:ham}) alone cannot stabilize
the ER equations, consider as fundamental variables not $\lr$ and
$\lt$, but instead the trace and the trace-free part of $L_{ij}$,
which in spherical symmetry are given by $L^i_i\equiv L^r_r+2\lt$ and
$L^{\rm TF}\equiv L^r_r-\lt$.  If one constructs evolution equations
for $L^i_i$ and $L^{\rm TF}$, one finds that perturbations of $L^{\rm
TF}$, holding $L^i_i$ and all other ER variables fixed,
obey~(\ref{padvective}) with
\begin{equation}
R(r) = {N(10\kt-7\krtil)\over 3} 
	        ={z^3(48+41z+34z^2+27z^3)\over 12M(1+z)^2(1+z^2)^2}.
		\label{fourrhs:ltracefree}
\end{equation}
The perturbations grow rapidly with time because $R(r)$ is large and
positive.  The source of the problem is a large, positive $L^{\rm TF}$
term on the right-hand side of the $L^{\rm TF}$ evolution equation.
Because~(\ref{erss:cons:ham}) involves only the trace of $L_{ij}$ and
not its trace-free part, this equation cannot be used to eliminate the
$L^{\rm TF}$ term and thus cannot be used to stabilize the system.

If one wishes to use the ER formulation in a 3D free evolution, one
must find a way of dealing with the unstable continuum mode afflicting
the ER evolution equations. Unfortunately, the above analysis suggests
that in 3D, this cannot be done in a simple way using algebraic
constraint equations. Accordingly, for 3D simulations it may be more
fruitful to pursue other hyperbolic formulations such as the
Einstein-Bianchi system, which, according to our analysis, should not
suffer from this type of instability.

\section{Acknowledgments}\label{sec:ack}
We thank Andrew Abrahams, \'Eanna Flanagan, and James York for helpful
discussions.  We also thank the anonymous referee for useful
suggestions. This work was supported by the NSF Binary Black Hole
Grand Challenge Grant Nos. NSF PHY 93-18152/ASC 93-18152 (ARPA
supplemented), NSF Grant PHY 94-08378 at Cornell, and NSF Grant AST
96-18524 and NASA Grant NAG 5-3420 at Illinois. Some computations were
performed on the Cornell Theory Center SP2 and on the National Center
for Supercomputing Applications SGI Origin 2000.

\appendix
\section{}\label{sec:append:a}
\subsection{Solution of Eq.~(\protect\ref{padvective}) on infinite domain}
\label{sec:appendix:soln}

Solutions to~(\ref{padvective}) propagate along characteristic curves
$r=r(t)$ that depend only on the shift vector and are defined by
\begin{equation}
{dr\over dt} = -\beta(r). 
\label{chareq}
\end{equation}
Each spacetime point $(r,t)$ intersects exactly one characteristic
curve. If we define $s(r,t)$ to be the radial coordinate at which the
characteristic curve passing through $(r,t)$ intersects the initial
slice $t=0$, then for $\beta(r)$ given by~(\ref{shid:beta}) we can
integrate~(\ref{chareq}) to find a relation between $s$, $t$, and $r$:
\begin{eqnarray}
{t\over 2M} = \ln{s\over r} 
 &+&{1\over 3}\left(r\over 2M\right)^3\left[\left(s\over r\right)^3-1\right]
\nonumber \\
 &+&{1\over 2}\left(r\over 2M\right)^2\left[\left(s\over r\right)^2-1\right]
\nonumber \\
 &+&{r\over 2M}\left[{s\over r}-1\right].
	   \label{charsol}
\end{eqnarray}

Treating $s$ and $t$ as independent variables, we can write
Eq.~(\ref{padvective}) in the form
\begin{equation}
\left.\partial \xi\over\partial t\right|_{s={\rm const}} = R(r(s,t))\xi.
\label{padvectivechar}
\end{equation}

Each value of $R(r)$ listed
in~(\ref{fourrhs}) can be written in the form
\begin{equation}
R(r) = {z^3(a+bz+cz^2+(b+c-a)z^3)\over 2M(1+z)^2(1+z^2)^2},\label{rgeneral}
\end{equation}
where $a$, $b$, and $c$ are constants and $z\equiv 2M/r$.  Using this
expression for $R(r)$ we can integrate~(\ref{padvectivechar}) together
with~(\ref{charsol}) to obtain the general solution
\begin{eqnarray}
\xi(r,t) = \xi_0(s)
	   \left(s\over r\right)^a
	   &&
	   \left(1+2M/r\over 1+2M/s\right)^{b-a}
		\nonumber\\&&\times
	   \left(1+(2M/r)^2\over 1+(2M/s)^2\right)^{(c-a)/2},
	   \label{pertsolt}
\end{eqnarray}
where $\xi_0(r)$ denotes $\xi$ on the initial slice $t=0$.

For a fixed value of $r$ we have $s\gg r$ at late times,
so~(\ref{charsol}) reduces to $t\sim s^3/12M^2$ and~(\ref{pertsolt})
reduces to
\begin{equation}
\xi(r,t) \sim \xi_0(12^{1/3}M^{2/3}t^{1/3})\left(12M^2t\over r^3\right)^{a/3},
	   \label{pertsollarget}
\end{equation}
where time-independent factors have been dropped. If $\xi$ initially
falls off like $r^{-m}$, then for a fixed $r$ it will behave like
$t^{(a-m)/3}$ at late times.  For $a>m+3$, perturbations will grow
superlinearly with time, but for $a\le m+3$ the growth is at most
linear (for $a<m$ the perturbation is actually damped), so it does not
represent an instability.

For the $\lr$ equation ($a=10$) to be individually stable, numerical
errors must fall off at least as fast as $r^{-7}$.  For the $\kt$
equation ($a=4$) to be stable, the leading-order falloff rate must be
no slower than $r^{-1}$. The $\mrt$ and $\at$ equations ($a=2$) will
be stable even if numerical errors grow with radius, as long as these
errors grow no faster than $r$.

Empirically, we find that the dominant numerical errors in the
wavelike variables ($\lr$, $\lt$, $\mrt$, $\arr$, and $\aor$) fall off
like $r^{-1}$ and propagate outward from the strong-field region near
the hole. This is what one would expect for modes that behave like
gravitational radiation (these modes are not allowed in spherical
symmetry but nevertheless can be present in numerical error terms).
The dominant errors in other variables also propagate outward from the
strong-field region, and fall off either like $r^{-1}$ or $r^{-2}$.
These falloff rates explain our observation that the $\lr$ equation is
individually unstable but the $\kt$, $\mrt$, and $\at$ equations are
individually stable.

For background solutions other than~(\ref{shid}), the forms of
$\beta(r)$ and $R(r)$ will be different, so the details of the
solution~(\ref{pertsolt}) will change. For example, if one takes the
MMIEF solution as a background (this is not relevant for
Eqs.~(\ref{erss:ev}) because the MMIEF solution is not preserved under
harmonic slicing, but is relevant for other systems of evolution
equations to which one might apply this analysis), $R(r)$ typically
falls off like $r^{-2}$ instead of $r^{-3}$, and $\beta(r)$ falls off
like $r^{-1}$ instead of $r^{-2}$, so the stability criterion becomes
$a\le m+2$ instead of $a\le m+3$.  At the same time, the coefficient
$a$ is typically smaller for the MMIEF background, so both the MMIEF
background and the background~(\ref{shid}) yield similar predictions for
stability.

Furthermore, note that our stability criterion can be applied to {\it
coupled\/} evolution equations as long as there are no couplings
through derivatives. For example, consider the coupled system
consisting of all ER variables except $\mrr$, $\mrt$, and $\aor$. If
$\mrr$, $\mrt$, and $\aor$ are held fixed, the perturbation equation
for the thirteen other variables can be written in the
form~(\ref{padvective}), where in this case $\xi$ is a
thirteen-element column vector and $R(r)$ is a $13\times 13$ matrix.
To determine stability, one examines each eigenmode of the
perturbation equation in the manner described above.  An example in
which this analysis cannot be used without modification is the coupled
system consisting of $\lr$ and $\mrr$.  In this case, the spatial
derivatives of $\lr$ in the $\mrr$ equation~(\ref{erss:ev:mrr}) and
the spatial derivatives of $\mrr$ in the $\lr$
equation~(\ref{erss:ev:lrr}) prevent one from writing down a matrix
perturbation equation of the form~(\ref{padvective}). Instead, the
perturbation equations possess more than one family of characteristic
curves, so the solution is more complicated.

\subsection{Solution of Eq.~(\protect\ref{padvective}) on finite domain}
\label{sec:appendix:solnfinite}
In numerical simulations one often does not have a domain that extends
to $r=\infty$, but instead one imposes an artificial boundary
condition at some finite value of $r$.  For simplicity, consider a
Dirichlet condition: assume $\xi$ is fixed to some constant value
$\xi_0$ at the outer boundary $r=r_0$.  If we let $t_0(r)$ be the time
it takes for information to propagate from the outer boundary $r_0$ to
some radius $r<r_0$, then for $(t,r)$ such that $t>t_0(r)$, the
solution of~(\ref{padvective}) is time-independent, and for $(t,r)$
such that $t<t_0(r)$, the solution is the same as for the case
considered in Appendix~\ref{sec:appendix:soln}.

For $\beta(r)$ given by~(\ref{shid:beta}), the time it takes for
information to propagate inward from radius $s$ to radius $r<s$ is
given by~(\ref{charsol}).  In this case, for $R(r)$ given
by~(\ref{rgeneral}) the time-independent solution is
\begin{eqnarray}
\xi(r) = \xi_0
	 \left(r_0\over r\right)^a
	   &&
	 \left(1+2M/r\over 1+2M/r_0\right)^{b-a}
		\nonumber\\&&\times
         \left(1+(2M/r)^2\over 1+(2M/r_0)^2\right)^{(c-a)/2}.
	 \label{pertsolnot}
\end{eqnarray}
For $r_0\gg r$, the time-independent solution behaves roughly like
$r^{-a}$.

One consequence of the above analysis is that if one uses a Dirichlet
outer boundary condition and an unstable mode of this type is present
(that is, if numerical perturbations fall off more slowly than
$r^{-a}$), then the instability will become more severe if the outer
boundary location is moved to a larger radius. This is because the
unstable mode has more time to grow before the time-independent state
is reached.  We have verified this numerically for the simple case of
the $\lr$ evolution equation solved with all other variables held
constant.

% REFERENCES
%###\bibliographystyle{prsty}
%###\bibliography{ref}

\end{document}